\documentclass[12pt]{article}
\usepackage[utf8]{inputenc}
\usepackage{fancyhdr}
\usepackage{url}
\usepackage[shortcuts]{extdash}
\usepackage{todonotes}
\usepackage{hyperref}
\usepackage{multirow,booktabs,subcaption,amsfonts,dcolumn}
\newcolumntype{d}[1]{D..{#1}}

 {\fancyhead{}
\fancyfoot[c]{\small{\rule{17.5cm}{1pt}}}}

\begin{document}
\title{\vspace{-2.5cm}
\begin{center}
\textbf{\small{SIGIR WORKSHOP REPORT}}\\\vspace{-0.5cm} \rule{17.5cm}{1pt}
\end{center}
\vspace{1cm}\textbf{Report on the 3rd Joint Workshop on Bibliometric-enhanced Information Retrieval and Natural Language Processing for\\ Digital Libraries (BIRNDL 2018)}}

\author{
Philipp Mayr\\
GESIS -- Leibniz Institute for the Social Sciences, Germany\\
\emph{philipp.mayr@gesis.org}
\and
Muthu Kumar Chandrasekaran\\
NUS School of Computing, Singapore\\
\emph{cmkumar087@gmail.com}
\and
Kokil Jaidka\\
University of Pennsylvania, USA\\
\emph{jaidka@sas.upenn.edu}
}

\date{}

\maketitle \thispagestyle{fancy} 

\abstract{The $3^{rd}$ joint BIRNDL workshop was held at the 41st ACM SIGIR Conference on Research and Development in Information Retrieval (SIGIR 2018) in Ann Arbor, USA. BIRNDL 2018 intended to stimulate IR researchers and digital library professionals to elaborate on new approaches in natural language processing, information retrieval, scientometrics, and recommendation techniques that can advance the state-of-the-art in scholarly document understanding, analysis, and retrieval at scale. The workshop incorporated three paper sessions and the $4^{th}$ edition of the CL-SciSumm Shared Task.} 


\section{Introduction}
The goal of the BIRNDL workshop at SIGIR is to engage the information retrieval (IR), natural language processing (NLP), digital libraries, bibliometrics and scientometrics communities to advance the state-of-the-art in scholarly document understanding, analysis and search and retrieval at scale \cite{mayr_report_2017}.
Scholarly documents are indexed by large, cross-domain digital repositories 
such as the ACL Anthology, ArXiv, ACM Digital Library, IEEE database, Web of Science, Google Scholar and Semantic Scholar. Currently, digital libraries collect and allow access to papers and their metadata --- including citations --- but mostly do not analyze the items they index. 
The  scale of scholarly publications poses a challenge for scholars in their search for relevant literature. Information seeking and sensemaking from the large body of scholarly literature is the key theme of BIRNDL and sets the agenda for tools and approaches to be discussed and evaluated at the workshop.

Papers at the $3^{rd}$ BIRNDL workshop incorporate insights from IR, 
bibliometrics and NLP to develop new techniques to address the open problems such 
as evidence-based searching, measurement of research quality, relevance and impact, 
the emergence and decline of research problems, identification of scholarly relationships and influences and applied problems such as language translation, question-answering and summarization. 
We also address the need for established, standardized baselines, evaluation metrics and test collections. Towards the purpose of evaluating tools and technologies developed for digital libraries, we are organizing the $4^{th}$ CL-SciSumm Shared Task based on the CL-SciSumm corpus, which comprises over 500 computational linguistics (CL) research papers, interlinked through a citation network.

\section{Overview of the papers}

This year six full papers were submitted to the workshop, three
of which were finally accepted as full papers for presentation and inclusion in the proceedings. In addition three poster papers were accepted for inclusion in the proceedings. 
The workshop featured one keynote talk, one full paper session, one session with presentations of systems participating in the CL-SciSumm Shared Task (see the CL-SciSumm overview paper \cite{Jaidka18}) and a poster session.
The following section briefly describes the keynote and sessions. 

\subsection{Keynote}
Byron Wallace gave an inspiring keynote on ``Automating Biomedical Evidence Synthesis: Recent Work and Directions Forward''~\cite{Wallace18}. He talked about recent progress in biomedical evidence synthesis and called on the NLP, IR and machine learning communities to take up the challenges that remain unaddressed in this critical field. He said that the field puts forth technically challenging problems of interest such as, building models with low-supervision, joint inference and extraction over long documents and hybrid crowd-expert annotations, to the aforementioned technical communities.

\subsection{Research papers}

Alzogbi \cite{Alzogbi18} presented a time-aware recommender system (Time-aware Collaborative Topic Regression - T-CTR) that accounts for the concept-drift in user interest by computing user-specific concept drift scores. The paper considered the use case of scientific papers recommendation and conducted experiments on data from citeulike. Results showed the superiority of the time-aware recommendation system T-CTR over the state-of-the-art systems.

Shinoda and Aizawa \cite{Shinoda18} proposed an unsupervised query-based summarization of scientific papers.
Importance scores for words calculated from word embeddings trained on an auxiliary corpus are used to compute sentence vectors. Finally, a random walk is performed on sentences which leverages distributional similarities between query terms and words in the sentence, as well as the similarities between pairs of sentences.

Brochier et al. \cite{Brochier18} applied a new document-query methodology to evaluate experts retrieval from a set of queries sampled directly from the experts documents. They provided a formal definition of the expert finding task and worked on a topic-query and a document-query evaluation protocol. They performed a detailed evaluation with three baseline expert recommender algorithms on two  AMiner expert data sets.

\subsection{Poster papers} 

Scharpf et al. \cite{Scharpf18} proposed a Wikidata markup to link semantic elements of a mathematical formula in MathML to Wikidata items. They suggested Formula Concept Discovery as a concept to develop automatic retrieval functions (e.g. formula clustering) on labeled full text corpora. They argued in favor of a larger MathML benchmark for evaluation purposes.

Jia and Saule \cite{jia18} proposed ``Keyphraser'' to alleviate the ``over-generation error'' when extracting key phrases from scientific documents. KeyPhraser is an unsupervised method for identifying document phrases using features such as concordance, popularity, informativeness and position of the first occurrence of the phrase.

Luan et al. \cite{yi2018} presented their SemEval-2018 Task 7 system Semantic Relation Extraction and Classification in Scientific Papers as an invited poster. They were invited due to the work's close relevance to the workshop and the authors' interest.\footnote{BIRNDL organising committee and \cite{yi2018} thank the BIRNDL reviewers for reviewing this accepted SemEval paper without prejudice.}

\subsection{CL-SciSumm}
We hosted the $4^{th}$ Computational Linguistics Scientific Summarization 
Shared Task, sponsored by Microsoft Research Asia as part of the BIRNDL 
workshop. The Shared Task is aimed at creation of an open corpus for citation 
based faceted summarization of scientific documents and evaluation of  
systems over three sub-tasks to output a summary. This is the first 
medium-scale shared task on scientific document summarization in the 
computational linguistics (CL) domain. The task and its corpus have the 
potential to spur further interest in related problems in scientific 
discourse mining, such as citation analysis, query-focused question 
answering and text reuse.

We have been incrementally building our annotated corpora over the four 
editions of CL-SciSumm. CL-SciSumm'18 systems were provided with 50 
document sets for training and evaluation was run over 10 test document 
sets. Eleven teams registered and ten teams participated in this year's shared task, on a corpus that was 33\% larger than the 2017 corpus. A total of 60 runs were evaluated for Task 1, and 33 runs were evaluated for Task 2.  For Task~1A, on using sentence overlap (F1 score) as the metric, the best performance was achieved by using voting methods on the results of a random forest classifier trained on lexical features and word embeddings, by NUDT \cite{system_6}. When ROUGE-based F1 is used as a metric, the best performance was also based on a BM25 voting mechanism by Klick Labs \cite{system_9}. The best performance in Task~1B was using kernel-based methods by CIST \cite{system_2}.

For Task~2, the team from LaSTUS/TALN+INCO had the best performance using CNN to learn the relation between a sentence and a scoring value indicating its relevance, and choosing the most relevant sentences for the summary \cite{system_11}. 
\begin{table}
\begin{center}
\begin{tabular}[!ht]{ l| l l l}
\toprule
  & \multicolumn{3}{c}{Year}\\
 \cline{2-4}
 & 2016 & 2017 & 2018\\
 
 \midrule
 Corpus size & 30 & 40 & 60\\
    \multicolumn{4}{c}{Task 1A results}\\
   \midrule
F1-score &  0.13 & 0.12 & 0.14\\
\midrule                                             
\shortstack{\footnotesize Best performing\\ approach} & Tf-idf & \shortstack{Weighted\\ensemble} & CNN \\
\midrule
  \multicolumn{4}{c}{Task 1B results}\\
   \midrule
F-1 score &  0.65 & 0.40 & 0.38\\
 \midrule
    \multicolumn{4}{c}{Task 2 (Rouge-2 scores)}\\
   \midrule
vs. Abstracts &  0.68 & 0.35 & 0.33\\
vs. Humans &  0.22 & 0.27 & 0.25\\
vs. Community &  0.25 & 0.20 & 0.21\\
\bottomrule

\end{tabular}
\end{center}
\caption{\footnotesize Task 1A, 1B and Task 2 scores and best performing approaches over the years.}
\label{tab:table1}
\end{table}

Table \ref{tab:table1} summarizes the best results, together with information about the corpus size in the different years. It also mentions the best performing approaches for Task 1. We observe that although the corpus size has doubled over the years, there is no perceptible change in the best scores. It should be noted that because of the difference in corpus sizes, the numbers are perhaps not comparable -- model overfitting has surely decreased since the efforts in 2016. We believe that to observe a real spurt in progress, there is a need to try to automatically scale dataset development and achieve O(N) = 1000 for any significant nudge in results, and we invite other researchers to collaborate with us on this long term goal.

Based on the trends in system performance, we anticipate that lexical methods complemented with domain-specific word embeddings in a deep learning framework would be an ideal approach for a scientific summarization task such as ours. For a detailed breakdown of the results, please see the CL-SciSumm'18 overview paper \cite{Jaidka18}. It is inspiring is that each year, we have had new research teams participating, including participants from the industry in 2018. Many other teams have consistently participated over the years and patiently worked with us to improve the quality of our corpus, and we are glad to have their support as well as to support their research with our efforts.

\section{Outlook and further reading}
With this continuing workshop series we have built up a sequence of 
explorations, visions, results documented in scholarly discourse, and 
created a sustainable bridge between bibliometrics, IR and NLP. We see 
the community still growing.

We will continue to organize similar workshops at IR, DL, Scientometric, NLP 
and CL premier venues. The combination of research paper presentations, and 
a shared task like CL-SciSumm with system evaluation has proven to be a 
successful agile format. We propose to continue with this format. 
CL-SciSumm '19 will see a major expansion in the training data size.  

In \looseness=-1  2015 we published a first special issue on ``Combining Bibliometrics and Information Retrieval'' in the \emph{Scientometrics} journal \cite{MayrS15a}.  Recently, a special issue on ``Bibliometrics, Information Retrieval and Natural Language Processing in Digital Libraries'' appeared in September 2018 in the \emph{International Journal on Digital Libraries}, see an overview in \cite{BIRNDL-SP-IJDL}. Another special issue on ``Bibliometric-enhanced Information Retrieval and Scientometrics'' appeared in \emph{Scientometrics}, see editorial \cite{BIR2018-SI}.
Since 2016 we maintain the ``Bibliometric-enhanced-IR Bibliography''\footnote{\url{https://github.com/PhilippMayr/Bibliometric-enhanced-IR_Bibliography/}} that collects scientific papers which appear in collaboration with the BIR/BIRNDL organizers. We invite interested researchers to join this project and contribute related publications.

\section*{Acknowledgments}

We thank Microsoft Research Asia for their generous support in funding the 
development, dissemination and organization of the CL-SciSumm dataset and 
the Shared Task\footnote{\url{http://wing.comp.nus.edu.sg/cl-scisumm2018/}}.
We are also grateful to the co-organizers of the $1^{st}$ BIRNDL 
workshop - Guillaume Cabanac, Ingo Frommholz, Min-Yen Kan and Dietmar 
Wolfram, for their continued support and involvement.
Finally we thank our programme committee members who did an excellent reviewing job. All PC members are documented on the BIRNDL website\footnote{\url{http://wing.comp.nus.edu.sg/birndl-sigir2018/}}.

\bibliographystyle{abbrv} 
\bibliography{sigirForumBirndl2018}
\end{document}